\documentclass{article}
\usepackage{arxiv}

\usepackage{booktabs}                               
\usepackage{tabularx}                       
\usepackage{soul}
\usepackage[utf8]{inputenc} 
\usepackage[T1]{fontenc}    
\usepackage{hyperref}       
\usepackage{url}            
\usepackage{booktabs}       
\usepackage{amsfonts}       
\usepackage{nicefrac}       
\usepackage{microtype}      
\usepackage[capitalise,sort&compress]{cleveref}    
\usepackage{lipsum}         
\usepackage{graphicx}
\usepackage{natbib}
\usepackage{doi}

\renewcommand{\cite}[1]{\citep{#1}}

\title{A Controlled Experiment on the Impact of Intrusion Detection False Alarm Rate on Analyst Performance}

\renewcommand{\shorttitle}{Impact of IDS False Alarm Rate}

\author{ \href{https://orcid.org/0000-0002-2534-8762}{\includegraphics[scale=0.06]{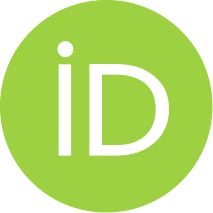}\hspace{1mm}Lucas Layman} \\
	Department of Computer Science\\
	University of North Carolina Wilmington\\
	Wilmington, NC 28403 \\
	\texttt{laymanl@uncw.edu} \\
	\And
	William Roden \\
	Department of Computer Science\\
	University of North Carolina Wilmington\\
	Wilmington, NC 28403\\
	\texttt{will.roden@ncino.com} \\
}

\hypersetup{
	pdftitle={A Controlled Experiment on the Impact of Intrusion Detection False Alarm Rate on Analyst Performance},
	pdfauthor={Lucas Layman, William Roden},
	pdfkeywords={cybersecurity, false alarms},
}

\begin{document}
	
	\maketitle
	
	\begin{abstract}
		Organizations use intrusion detection systems (IDSes) to identify harmful activity among millions of computer network events. Cybersecurity analysts review IDS alarms to verify whether malicious activity occurred and to take remedial action. However, IDS systems exhibit high false alarm rates. This study examines the impact of IDS false alarm rate on human analyst sensitivity (probability of detection), precision (positive predictive value), and time on task when evaluating IDS alarms. A controlled experiment was conducted with participants divided into two treatment groups, 50\% IDS false alarm rate and 86\% false alarm rate, who classified whether simulated IDS alarms were true or false alarms. Results show statistically significant differences in precision and time on task. The median values for the 86\% false alarm rate group were 47\% lower precision and 40\% slower time on task than the 50\% false alarm rate group. No significant difference in analyst sensitivity was observed.
	\end{abstract}
	
	\section{Introduction}
	Computer network defense is a collaboration between human analysts and automated systems. \textit{Intrusion detection systems} (IDSes) analyze network and system behavior and raise alarms to be investigated by security analysts. The analysts \textit{triage} the alarm to decide if it warrants further investigation \cite{DAmico2005}. The data generated by IDSes can be intractable and contributes to security analyst burnout~\cite{Durst1999,ChandranSundaramurthy2015}. \textit{False alarms} are the source of much of this excess, and as many as 99\% of IDS alarms are false alarms~\cite{Julisch2003}. Escalated IDS alarms entail a non-trivial investigation cost~\cite{Ryu2008}, contributing to fatigue and frustration among security analysts~\cite{Dykstra2018}.
	
	Ergonomics research has shown the negative consequences of false alarms in human-machine pairings~\cite{Bliss1995a, Meyer2001}. False alarms may lead to distrust in a system~\cite{Rice2009}, and an overabundance of alarms may lead an analyst to make incorrect decisions.  	Researchers observe that IDSes generate too much data for analysts to parse and argue that high IDS false alarm rates hurt analyst performance \cite{DAmico2005, Axelsson2000,Julisch2003}. However, these claims have not been investigated empirically.

%


	The frequency of alarms depends on how the IDS is configured, i.e., which rules are set to trigger an alarm. In practice, most of the alarms raised by IDSes are \textit{false alarms}; typical IDS false alarm rates are above 90\% with many as high as 99\% \cite{Julisch2003, Manganaris2000}. \citet{Axelsson2000} raises the issue of the \textit{base rate fallacy}, stating that the ratio of actual attacks to benign traffic is so low that IDSes must be extraordinarily accurate to have acceptable detection performance. 
	
	Operators' perceptions of alarm system reliability impacts their behavior. Controlled experiments show that operators who are told that an alarm system is correct 25\%, 50\%, or 75\% of the time will comply with the system's alarms at a commensurate rate~\cite{Bliss2003, Gerard2010, Guenzler2013}. Operators will also investigate or second-guess alarms more frequently when they know the system is unreliable~\cite{Gerard2010, Guenzler2013}. These findings hold whether the application is safety-critical or not. Further, studies show that unreliable systems yield slower response times to the alarms as operators must spend time verifying the accuracy of an alarm~\cite{Bliss1995a, Getty1995}. If the system does not provide enough information to confirm the accuracy of the alarm, operators tend toward the extremes of either complying with or ignoring all alarms in order to perform efficiently rather than accurately~\cite{Bliss2003, Wiczorek2010}. 

	A few studies have evaluated human analyst performance in cybersecurity tasks experimentally. \citet{Sawyer2018} examined how phishing detection accuracy and response time changed with varying signal probabilities.
	Their results suggest that as the ratio of malicious emails to benign emails becomes lower, the ability to detect malicious emails decays at a logarithmic rate. \citet{Ben-Asher2015} studied the role of domain knowledge in IDS alarm triage using a simulated IDS.	Their results showed that while general expertise is helpful in attack detection, situational knowledge of the specific network was more impactful on decision accuracy.
	\citet{dutt2012modeling} conducted a controlled experiment wherein participants evaluated simulated IDS alarms to determine if the alarm represented a threat. Their results showed that participants trained in the higher base rate (52\%) had higher hit rates and lower false alarm rates than participants trained in the lower base rate (12\%) when identifying a set of alarms with a 32\% true alarm rate.

	\subsection{Hypotheses}
	
	Identifying the relationship between analyst performance and IDS false alarm rate can help set benchmarks for IDS performance and help organizations understand the trade-off between high alarm rates and the ability to detect actual attacks. 
		A \textit{true alarm} is an alarm from the IDS that represents a malicious event, whereas a \textit{false alarm} is an alarm from the IDS that represents a benign event. The independent variable is the \textit{false alarm rate}: the ratio of false alarms to all alarms.
			
This study examines the following hypotheses: 
\begin{itemize}
	\item\textbf{H1:} Analyst \textit{sensitivity} (a.k.a. \textit{probability of detection}) decreases as the IDS false alarm rate increases; 
	\item\textbf{H2:} Analyst \textit{precision} (a.k.a. \textit{positive predictive value}) decreases as the IDS false alarm rate increases; 
	\item\textbf{H3:} Analyst \textit{time on task} (total time taken to triage all alarms) increases as the IDS false alarm rate increases.
	\end{itemize}

\section{Method}

\subsection{Participants}

Fifty-two (52) voluntary participants were recruited via computing-related major email lists, the Information Technology security office, and from two computer security courses in the Fall of 2019 at the University of North Carolina Wilmington. Participants were entered into a gift card raffle or given course extra credit as compensation regardless of performance.

\subsection{Materials}

Participants were presented with the following scenario: ''You are a junior cyber security analyst at Company XYZ. Your job is to perform initial \textit{triage} on a list of network events that your system's IDS has deemed an alert. You must determine if the alert could be an attack against the network, requiring further investigation by a senior analyst, or is normal network activity that can be dismissed.'' 

The simulated IDS alarms shown in the experiment were based on \textit{impossible travel} scenarios where a user authenticates from two geographic locations within a time period where physical travel between the locations would be impossible. 
The impossible travel scenario was chosen because of its intuitiveness and because it is a rule actively investigating by real IDS systems. 
The dataset and its evaluation are reported in~\citet{Roden2020}. The dataset contains 30 true alarms and 43 false alarms.
An example of an \textit{impossible travel} alarm is shown in \cref{tab:impossible}. False alarms in the data set are due to the use of a virtual private network (VPN), a mobile device, a cloud server, or borderline realistic travel times. 
\begin{table}[ht!]
\centering
\caption{Event \#66 --- A True Alarm with Password Guessing}
\label{tab:impossible}
\begin{tabular}{@{}lrr@{}}
	\toprule
	City of Authentication   & Seattle     & Moscow     \\ 
	\# Successful Logins     & 4           & 11         \\
	\# Failed Logins         & 1           & 3          \\
	Source Provider          & Telecom     & Telecom    \\ \midrule
	Time between Authentications & \multicolumn{2}{c}{1.75} \\
	VPN Confidence           & \multicolumn{2}{c}{0\%}  \\ \bottomrule
\end{tabular}
\end{table}

Each alarm contains the following data:
\begin{enumerate}
\item \textbf{Cities of Authentication}: the two geographic locations from which the IDS detected an authentication. 
\item \textbf{Number of Successful Logins}: the number of successful authentications from each location in the past 24 hours.
\item \textbf{Number of Failed Logins}: the number of failed logins from each location in the past 24 hours.
\item \textbf{Source Provider}: the type of internet provider the authorizations came from at each location. Possible values are: telecom internet provider, mobile/cellular, and cloud host.
\item \textbf{Time between Authentications}: the shortest time in decimal hours between authentications from the two cities in the past 24 hours.  This is the field that triggers an alarm in a real IDS.
\item \textbf{VPN Confidence}: a percent likelihood that the user utilized a VPN.
\end{enumerate}

The experiment was implemented in a custom web application~\cite{Roden2019a} similar to that of \citet{Ben-Asher2015}. 
The web application displays sample alarms for training, a "Security Playbook" containing guidance on how to evaluate IDS alarms, and the main task of evaluating 52 alarms.  For each training and main task alarm, participants select "Escalate" 
if the alarm should be sent to a senior analyst for investigation or "Don't escalate" if the participant believes the alarm is a false alarm. Participants rate their level of confidence in their decision on a five point scale. Participants can also choose "I don't know" for their decision in which case a confidence score is not required. The sample alarm training pages include a button to reveal the correct answer and supporting rationale. The "Security Playbook" contains heuristics for evaluating an alarm, a concern level (Low, Medium, High) for each geographic location in the dataset, and a table of typical travel times between the locations.  This information enables participants to discern false alarms. The playbook can be viewed at any time in a pop-up window.

\subsection{Procedure}

 
Four separate sessions were conducted with different participants in on-campus computer labs with dual monitors. Each session was proctored by the second author and lasted approximately 60 minutes. Each session consisted of: 
\begin{enumerate}
	\item completing informed consent and a background information form;
	\item training on the simulated IDS environment;
	\item the main task of IDS alarm evaluation;
	\item completing a post-questionnaire including the NASA Task Load Index (TLX)~\cite{Hart1988} to evaluate task workload.
\end{enumerate}
All participants completed the experiment in under an hour.

Participants were randomly assigned to two treatment groups and had no knowledge their treatment group. The 50\% False Alarm Rate (FAR) group saw 25 \textit{true alarms} ("Escalate" is correct) and 25 \textit{false alarms} ("Don't escalate" is correct), while the 86\% FAR group saw seven \textit{true alarms} and 43 \textit{false alarms}. The false alarm rate of 50\% is conservatively the upper threshold a human operator can tolerate before performance suffers~\cite{Axelsson2000}, while 86\% is closer to the false alarm rate of actual IDS systems~\cite{Julisch2003}. The alarms for each treatment group were randomly sampled without replacement from the alarm data set. All participants within a group were presented with the same set of alarms in random order. 

In the training phase, participants were introduced to the Security Playbook and evaluated five training alarms. The first two training alarms (one true and one false) presented analysis of the alarm information and rationale for the correct decision. The remaining three training alarms (two false and one true) were presented as in the main task except that participants could reveal the correct answer and its rationale. Training alarm evaluations were not analyzed.


In the main task, participants were shown a table of links to the 52 alarms and a banner indicating how many alarms remained to evaluate. The table also displayed the participant's recorded decision for each alarm. 
Each alarm page displayed a table of information similar to \cref{tab:impossible} and a form to record the "Escalate/Don't Escalate/I don't know" decision and confidence rating. Once a decision was made, participants were automatically shown the next alarm in the table. Participants could use the table of links to visit the alarms in any sequence to review past decisions or to change their past responses if their evaluation schemas updated during the main task. Of the 52 participants and 2704 final alarm evaluations, only four participants changed their decision or confidence on a total of 10 alarms (0.4\%). To help identify participants who were simply "clicking through" the main task to earn credit, two of the 52 alarms were "verification alarms" whose content gave specific instructions for which decision and confidence to select.

Participant decisions and confidence ratings for each alarm were recorded by the app. The time on task for each participant is measured from when the participants clicked a "Begin Experiment" button to initiate the main task to when they clicked a "Complete Experiment" button on the main task list page.

Participants were neither informed of the reliability of the IDS nor given feedback on the accuracy of their evaluations. The IDS's reliability had to be estimated through comparing alarms to the Security Playbook. This lack of feedback reflects real cybersecurity analysis.

\section{Results}
\label{sec:analysis}




Fifty-two (52) participants completed the experiment. One participant's data from the 50\% FAR group was discarded because they responded incorrectly to a verification alarm. 
The following analysis includes 25 participants in the 50\% FAR group and 26 participants in the 86\% FAR group. The data is not normally distributed, and the two-tailed Mann-Whitney test is used for hypotheses testing. Effect sizes ($r_U$) are computed as the rank-biserial coefficient of correlation \cite{Wendt1972}. The experiment data and analysis package are published online \cite{Roden2019b}.


The performance statistics are presented in \cref{tab:performance_treatment}, and box-and-whisker charts of the data are presented in Figures~\ref{fig:sensitivity}--\ref{fig:time-on-task}. Of the 2704 final alarm decisions, 25 (0.9\%) "I don't know" decisions were recorded and are excluded from analysis.
The mean \textit{confidence} for the alarm decisions was 3.84 for both groups ($U$ = 331.0, $p$  = .917, $r_U$ = .019
). 
The post-experiment NASA TLX scores across groups are shown in \cref{tab:tlx}. No significant differences were found between groups for any dimension; a finding also of \citet{Sawyer2018}.  



\begin{table}[!htb]
	
	\caption{Performance measures. "FAR" = False Alarm Rate.}
	\centering
	\begin{tabular}{llrr}
		\toprule
		& group &  50\% FAR &  86\% FAR \\
		& \textit{n}     &   25      & 26 \\
		\midrule
		
		sensitivity & mean &      0.78 &      0.69 \\
		& median &      0.83 &      0.86 \\
		& $\sigma$ &      0.19 &      0.25 \\
		& min &      0.31 &      0.29 \\
		& max &      1.00 &      1.00 \\ \midrule
		precision & mean &      0.80 &      0.42 \\
		& median &      0.80 &      0.33 \\
		& $\sigma$ &      0.12 &      0.27 \\
		& min &      0.58 &      0.09 \\
		& max &      1.00 &      1.00 \\ \midrule
		time on task & mean &     13.95 &     19.25 \\
		& median &     13.44 &     18.76 \\
		& $\sigma$ &      3.50 &      5.61 \\
		& min &     8.00 &     11.15 \\
		& max &     21.79 &     31.14 \\ 
			\midrule
			correctness & mean &      0.75 &      0.76 \\
			& median &      0.74 &      0.76 \\
			& $\sigma$ &      0.12 &      0.16 \\
			& min &      0.50 &      0.48 \\
			& max &      0.92 &      0.98 \\
		\bottomrule
	\end{tabular}

	\label{tab:performance_treatment}
\end{table}


\begin{table}[!htb]
	\centering
	\caption{NASA TLX scores. Each dimension is on a 10-point scale.}
	\begin{tabular}{@{}lrrrrrr@{}}
		\toprule 
		& \multicolumn{2}{c}{50\% FAR}     & \multicolumn{2}{c}{86\% FAR}      & \multicolumn{2}{c}{All}     \\ 
		              & med   & mean & med   & mean & med  & mean \\ \midrule 
		mental      & 6        & 5.6  & 6        & 5.7  & 6       & 5.7  \\
		physical    & 1        & 1.8  & 1        & 1.2  & 1       & 1.5  \\
		temporal    & 3        & 3.5  & 3        & 3.1  & 3       & 3.4  \\
		performance & 5        & 5.3  & 5        & 4.7  & 5       & 4.9  \\
		effort      & 5        & 5.1  & 4.5        & 4.7  & 5       & 4.9  \\
		frustration & 3        & 3.8  & 3        & 3.0  & 3       & 3.4 \\
		\bottomrule
	\end{tabular}
	\label{tab:tlx}
\end{table}

\subsection{Sensitivity (Probability of Detection)}

\begin{figure}[!htb]	
	\caption{Sensitivity (Probability of Detection)}
	\centering
	\includegraphics[width=0.45\textwidth]{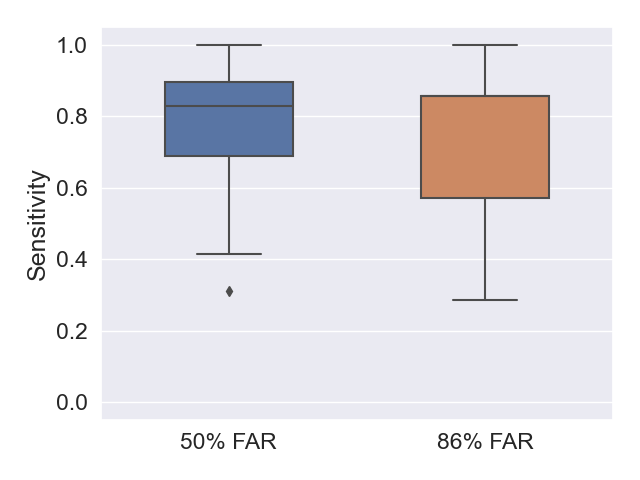}
	\label{fig:sensitivity}
\end{figure}

\textit{Sensitivity} is the ratio of true alarms escalated to all true alarms, i.e., the percentage of all true alarms escalated by the analyst. The median sensitivity of both groups was around 85\%, which indicates a good probability of detecting a true alarm, and the rank difference is not statistically significant ($U$ = 412.5, $p$  = .098, $r_U$ = .269
). The spread of both groups in \cref{fig:sensitivity} suggests meaningful variance in individual analysts' abilities to recognize true alarms in this task.

\subsection{Precision (Positive Predictive Value)}
\begin{figure}[!htb]
	\caption{Precision (Positive Predictive Value)}
	\centering
	\includegraphics[width=0.45\textwidth]{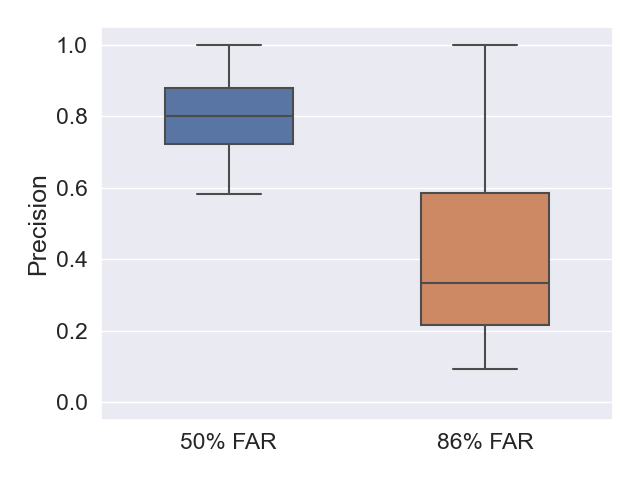}
	\label{fig:precision}
\end{figure}

\textit{Precision} is the ratio of escalated true alarms to all escalated alarms, i.e., the percentage of alarm escalations that were correct. The 50\% group's median precision was 0.80, while the 86\% group's precision was much lower at 0.33.
\cref{fig:precision} visualizes the precision between groups, and the rank difference is statistically significant ($U$ = 555.0, $p$  < .001, $r_U$ = .708
). Combined with the previous results, the data suggest that both groups perform well in hit detection, whereas the 86\% group tended to escalate more false alarms. In reality, escalating a false alarm is the safe option but incurs additional cost to investigate the alarm further.

\subsection{Time on Task}
\begin{figure}[!htb]
	\caption{Total Time on Task}
	\centering
	\includegraphics[width=0.45\textwidth]{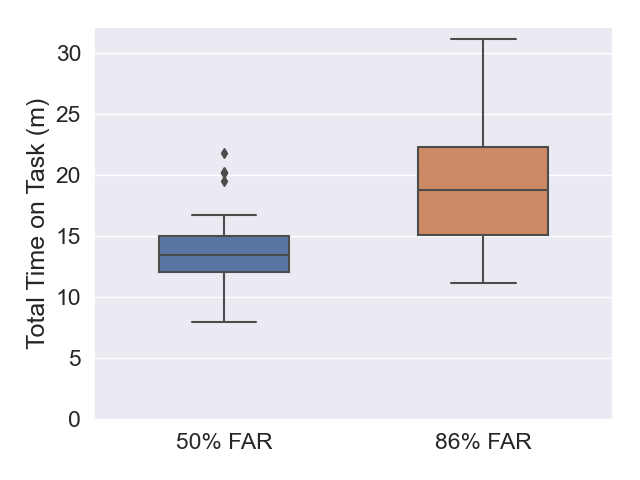}
	\label{fig:time-on-task}
\end{figure}

\begin{figure}[!htb]
		\caption{Mean Decision Time by Alarm Order in the Main Task}
		\centering
	\label{fig:event_decision_time}
	\includegraphics[scale=0.5]{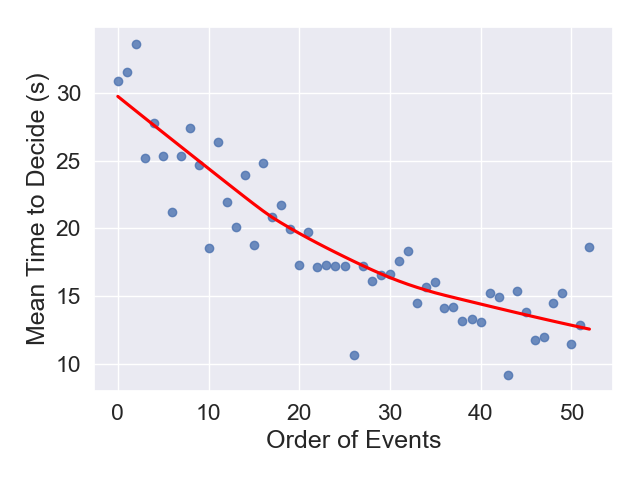}

\end{figure}

\cref{fig:time-on-task} shows that participants in the 86\% Group took approximately 40\% longer (median difference of 5.3 minutes) to complete the main task than the 50\% Group. The rank difference is statistically significant ($U$ = 143.0, $p$  < .001, $r_U$ = .560
). The mean time to evaluate a single alarm was 15.6s for the 50\% group and 21.4s for the 86\% group, and that rank difference is also statistically significant ($U$ = 147.0, $p$ < .001, $r_U$ = .548). \cref{fig:event_decision_time} shows that the mean time to make a decision decreased as the experiment progressed, suggesting that participants quickly developed schemas for evaluating alarms or that they became fatigued.

\section{Discussion}


The results do not support Hypothesis 1 that analyst sensitivity decreases as false alarm rate increases, a finding
somewhat contrary to \citet{Sawyer2018}. One explanation for the different findings is that Sawyer and Hancock observed their effect only at a 99\% false alarm rate. One interpretation is that significant differences in analyst sensitivity are more likely to occur when true alarms are rare. This would suggest that only severely overtuned IDS systems may decrease analyst's ability to identify true attacks.

The results support Hypothesis 2 that analyst precision decreases as false alarm rate increases. One explanation for this finding can be derived from \citet{dutt2012modeling}: the 50\% Group became more accustomed to seeing alarms that needed to be escalated, thus they were better able to determine when an alarm did not need to be escalated. Another explanation for the finding is best summarized by feedback from a participant in the 86\% Group, "I felt like there were only a few examples that seemed like they should definitely be escalated, but [I] escalated many more than that so that a threat would not go unnoticed." This mindset reflects the reality of cybersecurity practice: organizations place a priority on high sensitivity to intrusions at the expense of precision. The results suggest that an IDS with "only" a 50\% false alarm rate, a rate unacceptable in many disciplines, would yield significant improvements in analyst accuracy. 

The data support Hypothesis 3 that analyst time on task increases as false alarm rate increases. \citet{Getty1995} and \citet{Bliss1995a} both suggest that increased time is due to increased effort to verify rather than trust results when working with an unreliable system. Participants in our study were not informed of the system's reliability and had to form their own opinions based on the heuristics in the Security Playbook. Another explanation derived from \citet{dutt2012modeling} is that participants in the 86\% group did not see enough true alarms to develop an accurate schema for recognizing true alarms, thus requiring them to spend more time reasoning about potential true alarms. More analysis is needed to compare decision times on true alarm vs. false alarms between groups.

While we strove to make the task reflective of real cybersecurity practice, replicating a real IDS environment is challenging and access to security professionals is limited.
Performance on cyber analysis tasks is linked to domain knowledge~\cite{Ben-Asher2015}. However, our participants knowledge level may be closer to front-line IDS analysts per \citet{Zhong2018}, who state that many cybersecurity novices are tasked with triaging alarms. 

%

%
%

Our results suggest a few practical takeaways. 
First, a "reasonable" IDS false alarm rate of 50\% yielded a substantial increase in precision over the higher false alarm rate without a reduction in sensitivity. Such an improvement in the human-IDS team's ability to triage network activity accurately would reduce wasted investigation effort.
Second, the time taken by the 50\% group was on average 40\% less than the 86\% group while maintaining better performance, which would represent a substantial improvement in analyst efficiency if it scales. 
These results bear additional investigation with larger sample sizes, longer duration, varied alarm triage tasks, and populations that include a variety of cybersecurity expertise.

\bibliographystyle{unsrtnat}

\end{document}